\documentclass[aps,prb,twocolumn,showpacs,floats]{revtex4}
\usepackage{epsfig}
\usepackage{amsmath}
\newcommand{\be}{\begin{equation}}
\newcommand{\ee}{\end{equation}}
\newcommand{\bea}{\begin{eqnarray}}
\newcommand{\eea}{\end{eqnarray}}
\newcommand{\ba}{\begin{array}}
\newcommand{\ea}{\end{array}}
\newcommand{\nn}{\nonumber}

\newcommand{\Del}{\Delta}

\newcommand{\Gam}{\Gamma}

\newcommand{\al}{\alpha}

\newcommand{\ua}{\uparrow}
\newcommand{\da}{\downarrow}
\begin{document}

\title{\bf A spin pump turnstile: parametric pumping of a spin-polarized 
current through a nearly-closed quantum dot} 

\author{M. Blaauboer and C.M.L. Fricot
} 

\affiliation{Kavli Institute of NanoScience, Delft University of Technology, 
Lorentzweg 1, 2628 CJ Delft, The Netherlands}
\date{\today}

\begin{abstract}
We investigate parametric pumping of a spin-polarized current through a nearly-closed quantum dot in a perpendicular magnetic field. Pumping is achieved by tuning the tunnel couplings to the left and right lead - thereby operating the quantum dot as a turnstile - and changing either the magnetic field or a gate-voltage. We analyze the quantum dynamics of a pumping cycle and the limiting time scales for operating the quantum dot turnstile as a pure spin pump. The proposed device can be used as a fully controllable double-sided and bipolar spin filter and to inject spins ``on-demand''.
\end{abstract}

\maketitle

Parametric pumping of electrons in mesoscopic systems refers to the generation of a dc current by periodic modulations of two or more system parameters (e.g. a gate voltage or a magnetic field) in the absence of a bias voltage. Although the basic idea of transporting particles in this way dates back to 1983~\cite{thou83}, charge pumping only started to be actively investigated around a decade ago. The first experimental observation of a pumped current was obtained using a quantum dot in the Coulomb blockade regime~\cite{kouw92,poth91}. By alternately raising and lowering the tunnel barriers between the dot and two external leads, electrons were pumped one by one through the system resulting in a current that is quantized in units of $e \omega$, with $\omega$ the pumping frequency. This quantum dot turnstile acts as a "classical" pump, in the sense that it involves tunneling but no quantum interference. The latter does play an important role in the second pumping experiment, which was done using an open rather than a nearly-closed quantum dot~\cite{swit99,rectification}. Two oscillating gate voltages with a phase difference between them induce shape changes of the dot potential which, if applied sufficiently slowly (so that electrons can follow the motion of the potential adiabatically) cause charge to be moved across the dot. The resulting current is not quantized but depends on the microscopic properties of the system~\cite{brou98}. 

In recent years, attention has focussed on generating spin-polarized currents by parametric pumping of spin in semiconductors. An intruiging question in this context is whether a spin current can be pumped in the absence of a charge current, which is the case when equal amounts of electrons with opposite spins are displaced in opposite directions in the system. Several works have recently appeared that consider a variety of spin pumps~\cite{mucc02,wats03,gove03,aono03}. In the proposal by Mucciolo {\it et al.}~\cite{mucc02}, followed by the experiment in Ref.~\cite{wats03}, the spin and charge currents in an open quantum dot in a magnetic field are investigated and it was predicted and observed that under appropriate circumstances a spin current can be generated without net charge flow. This is the "spin analogue" of the charge pumping experiment in an open quantum dot~\cite{swit99}. Here we propose the ``spin analogue'' of the ``classical'' turnstile charge pump experiment~\cite{kouw92,poth91}. We consider two related but different turnstile models: in the first one the tunnel couplings to a left and a right lead and an oscillating magnetic field are used as pumping parameters. In the second model the latter is replaced by a time-dependent gate voltage. A similar model as the first one has recently been investigated by Aono {\it et al.}~\cite{aono03}, who studied pumping through a single orbital level in a quantum dot and predicted quantized amounts of spin-without-charge to be pumped in certain regions of parameter 
space. Here we ask the opposite question, namely how to {\it construct} the pumping mechanism so that it allows for reliable and fully controllable operation as a spin pump in the entire chosen parameter space. We analyze and estimate the different time scales that are important during the pumping process: the tunneling time, the times needed to adjust the pumping parameters and the time scale for spin-flip processes and predict that under realistic circumstances the quantum dot turnstile can produce a pure spin current that is quantized in units of 2 spins/$\tau$, with $\tau$ the duration of a pumping cycle.

The system we consider consists of a quantum dot in a perpendicular magnetic field $B$ that is weakly coupled to two external leads (Coulomb-blockade regime). The dot can either be empty, occupied by a single electron with spin-$\ua$ or spin-$\da$ or occupied by two electrons in the singlet state~\cite{fewelectrons}. The energy of each of these states is given by $E_0 = 0$, $E_{\ua}$, $E_{\da} = E_{\ua} + E_Z$ and $E_S = E_{\ua} + E_{\da} + E_C$, with $E_{\ua}$ the single-particle energy for a spin-$\ua$ electron, $E_Z \equiv |g^*| \mu_B B$ the Zeeman energy and $E_C = e^2/C$ the charging energy, where $C$ is the total capacitance of the dot. The corresponding chemical potentials are
\begin{subequations}
\bea
\mu_{\ua (\da)} & = & E_{\ua (\da)} - E_0 = E_{\ua(\da)} \\
\mu_{S \leftrightarrow \ua (\da)} & = & E_S - E_{\ua (\da)} = E_{\da (\ua)} + E_C 
\eea
\end{subequations}

\begin{figure}
\centerline{\epsfig{figure=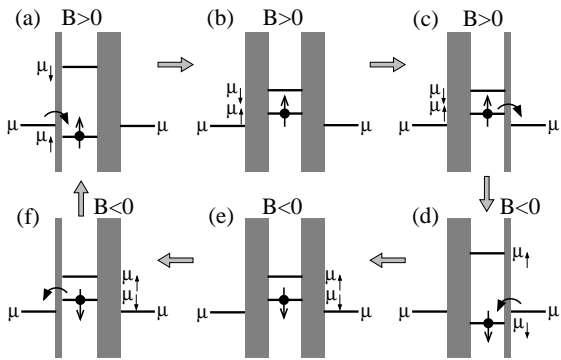,width=0.85\hsize}}
\caption[]{Schematic picture of a spin pumping cycle using an oscillating magnetic field. The levels represent the chemical potential of the dot when occupied by an electron with spin-$\ua$ ($\mu_{\ua}$) and spin-$\da$ ($\mu_{\da}$). $\mu$ denotes the chemical potential in the leads. See the explanation in the text.
}
\label{fig:cycle1}
\end{figure}

The first pumping scheme is schematically depicted in Fig.~\ref{fig:cycle1}. Going from (a) to (f) the pumping cycle consists of the following steps: the starting position is an empty dot with $\mu_{\ua} < \mu < \mu_{\da}$ and high tunnel barriers, so that the dot is well-isolated from the leads. Then the left barrier is lowered so that a spin-${\ua}$ electron can tunnel in from the left lead [(a)]. After that, this barrier is raised again to its starting position while simultaneously the magnetic field (and hence the Zeeman splitting) is slowly reduced [(b)]. When $\mu_{\ua} > \mu$, the right barrier is lowered so that the electron can tunnel out [(c)]. Meanwhile the magnetic field continues to be slowly decreased and becomes negative. In this reversed field $\mu_{\da} < \mu_{\ua}$ and when also $\mu_{\da} < \mu$ a spin-$\da$ electron tunnels into the dot from the right lead [(d)]. The right barrier is then closed and the magnetic field is slowly increased [(e)]. As soon as $\mu_{\da} > \mu$, the left gate is opened, so that the $\da$-spin can tunnel out into the left lead [(f)]. Raising the left barrier brings the system back to the beginning of the cycle with the net result that a spin-$\ua$ has been transferred from left to right, and a spin-$\da$ from right to left. Thus no net charge and 2 spins have been pumped across the dot.

In order for this spin-without-charge turnstile to operate succesfully and efficiently, a careful analysis of relevant time scales is required. Assuming that the raising and lowering times of the tunnel barriers are fast~\cite{raising}, we are left with three important time scales in the pumping cycle: the tunneling time $\tau_{\rm tunnel}$ for a spin to enter or leave the dot, the switching time $\tau_{\rm switch}$ of the magnetic field from $B$ to $-B$ or vice versa, and the spin flip time in the dot. In principle, the first two times should be as short as possible (in order to maximize the pumped current) and controlled in such a way to allow for reliable transfer of each spin (in order to maximize the spin current). The latter condition defines an upper bound on $\tau_{\rm switch}$: the field must be switched fast enough to avoid adiabatic transfer to another energylevel with the opposite spin. A straightforward way to fullfil this condition is by only starting to reverse the field after the electron has tunneled out of the dot. However, since tunneling takes relatively long, this might not be the most efficient way in view of minimizing $\tau$. Below, we therefore investigate the
probability of adiabatic transfer into another level during the switching of the magnetic field. To this end, consider the Hamiltonian that describes electrons confined to a quantum dot which is modeled by a parabolic potential,
\be
{\cal H} = \frac{(\vec{p} - e \vec{A})^2}{2m^{*}}
- \frac{1}{2} E_Z \sigma + \frac{1}{2} m^{*} \omega_0^2(x^2 + y^2).
\label{eq:Ham}
\ee
Here $m^{*}$ denotes the effective mass, $\sigma = \pm 1$ for spin-$\ua$ and spin-$\da$ resp. and we assume the magnetic field to be applied in the $z$-direction. 
By solving the Schr\"odinger equation for ${\cal H}$ one obtains the well-known Fock-Darwin energylevel
spectrum~\cite{Fock28}, 
\be
E_{n,l,\sigma}  =  (2n+|l| + 1) \hbar \omega - \frac{1}{2} l \hbar
\omega_c - \frac{1}{2} E_Z \sigma.
\label{eq:levels} 
\ee
Here $\omega$ $\equiv$ $\sqrt{\omega_0^2 + \omega_c^2/4}$ and $\omega_c$ = $eB/m^{*}$. The corresponding eigenfunctions for the lowest orbital levels with n=0 and l=0,$\pm$1 are given by $\psi_{0,0,\sigma}(r,\phi)$ = $A\, e^{-r^2/( 2 l_m^2)} \chi_{\sigma}$ and $\psi_{0,\pm 1,\sigma}(r,\phi)$ = $A\,(r/l_m)\, e^{-r^2/( 2 l_m^2)} e^{\pm i\phi}\, \chi_{\sigma}$, with $A \equiv$ $(\sqrt{\pi} l_m)^{-1}$, $l_m = (\hbar /(m^{*}\omega))^{1/2}$, $\chi_{1}= \left( \ba{l} 1 \\ 0 \ea \right)$ and $\chi_{-1}= \left( \ba{l} 0 \\ 1 \ea \right)$.

In the absence of
spin interactions such as spin-orbit or hyperfine interaction, adiabatic
transfer between the energylevels (\ref{eq:levels}) during the switching of the field is not allowed, since both $[{\cal H},S_z]=0$ and $[{\cal H},L_z]=0$. Any spin 
will thus remain in its eigenstate during the switching of the
field. This situation changes in the presence of Rashba spin-orbit (s.o.) interaction 
\be
{\cal H}_R = \frac{\al_R}{\hbar} [ \vec{\sigma} \times (\vec{p} - e \vec{A}) ]_z,
\label{eq:HamR}
\ee
with $\al_R$ the Rashba interaction constant. Since ${\cal H}_R$
commutes with $J_z$ but not with $S_z$, a spin-$\ua$ in the lowest orbital level could in principle be adiabatically transferred a higher orbital level for spin-$\da$. Applying (\ref{eq:HamR}) to the eigenfunctions $\psi_{n,l,\sigma}$ we obtain that the Rashba interaction couples $\psi_{0,0,1}$ 
to $\psi_{0,1,-1}$ and that the corresponding energylevels are modified as
\bea 
\tilde{E}_{0,0,1\, (0,1,-1)} & = & \frac{3}{2}\hbar \omega - \frac{1}{4}\hbar \omega_c \, - (+) \frac{1}{2} \left[ (\hbar \omega - \frac{1}{2} \hbar \omega_c + E_Z)^2 \right. \nn \\
& & \left. + \tilde{\al}_R (\hbar \omega - \frac{1}{2} \hbar \omega_c) \right]^\frac{1}{2}, 
\label{eq:Etilde}
\eea
with $\tilde{\al}_{R} \equiv 8 \al_R^2 m^{*}/\hbar^2$. From (\ref{eq:Etilde}) we see that the two levels do not cross during switching of the field and hence adiabatic transfer from the lowest spin-$\ua$ level to this higher-lying spin-$\da$ level does not occur. A similar reasoning applies for the lowest spin-$\da$ level. Therefore also in the presence of s.o. interaction adiabatic transfer of an electron in one of the Zeeman-split levels in Fig.~\ref{fig:cycle1} to any other level is not allowed. 

This is different, however, in the presence of  hyperfine interaction. Since the hyperfine Hamiltonian
\be
{\cal H}_{\rm hyp} = A S_z I_z + \frac{1}{2} A (S_{+} I_{-} + S_{+} I_{-}),
\label{eq:Hamhyp}
\ee
where $I$ represents the nuclear spin operator, commutes with $L_z$ but not with $S_z$, adiabatic transfer between the two lowest spin-split levels is in principle allowed. The time $T_{\rm ad}$ that is required for this process depends on the degree of polarization of the nuclei: if the nuclei are fully polarized along the direction of the magnetic field, $T_{\rm ad}$ for spin-$\ua$ to be transferred to spin-$\da$ is infinite, since simultaneous and opposite spin flips of the electron spin and a nuclear spin is not possible. For an arbitrary polarization of the nuclei we find, using (\ref{eq:Ham}), (\ref{eq:Hamhyp}) and following Ref.~\cite{mess58}
\be
T_{\rm ad} \gg \frac{\hbar N^2 |g^{*}| \mu_B B}{N_{\rm eff}^2A^2}
\label{eq:Tad}
\ee
for switching the field from $B$ to $-B$ as $B(t) = B[1 - 2t/T]$ for $0 \leq t \leq T$. Here $N_{\rm eff}/N$ is the fraction of the total number of nuclei $N$ in the dot which interact with the electron spin via ${\cal H}_{\rm hyp}$. 

From the above analysis we conclude that as long as $\tau_{\rm switch} \ll T_{\rm ad}$ and $\tau_{\rm switch} \leq \tau_{\rm tunnel}$, the total cycle time is given by $\tau = 4\, \tau_{\rm tunnel}$. For low temperatures the charge and spin currents are then defined as
\bea
I_c & = & \frac{e}{\tau}( n_{\ua,R} + n_{\da, R} - n_{\ua, L} - n_{\da, L} ) 
\\
I_s & = & \frac{1}{\tau}( n_{\ua,R} - n_{\da, R} - n_{\ua, L} + n_{\da, L} ).
\eea
Here $n_{\ua,R}$ denotes the number of spin-$\ua$ electrons that are transferred to the right during one pumping cycle etc.. In the absence of spin-flip scattering in the dot and other mechanisms such as co-tunneling that reduce the efficiency of the pumped spin current (see the discussion below) we find that $n_{\ua,R}=n_{\da, L}=1$ and $n_{\ua, L}=n_{\da, R}=0$, so that $I_c=0$ and $I_s = 2/\tau$. In the presence of spin-flip scattering, the number of transferred spins can be calculated by solving the three coupled Master equations
\be
\dot{\rho}_{i}  =  W_{ij}\rho_{j} + W_{ik}\rho_{k} - (W_{ji}
+ W_{ki}) \rho_{i} 
\label{eq:rhoeq}
\ee
for $(i,j,k)=(0,\ua,\da)$ and its two cyclic permutations.
Here $\rho_0$ and $\rho_{\ua (\da)}$ represent the probabilities that the dot is empty and occupied by a spin-$\ua$ (spin-$\da$) electron resp., and $W_{ij} \equiv W_{ij,R}+W_{ij,L}$ [for $(i,j)\neq (\ua,\da)$ or $(\da,\ua)$] is the transfer rate from state $j$ to state $i$ via tunneling from the left (L) and right (R) lead. $W_{\ua \da}$ and $W_{\da \ua}$ represent the spin-flip rates $W_{\ua \da}^{-1} = T_1 (1 + e^{-E_Z/k_B T})$ and $W_{\da \ua}^{-1} = T_1 (1 + e^{E_Z/k_B T})$, with $T_1$ the intrinsic spin-flip time at $T=0$. The solution of (\ref{eq:rhoeq}) is given by an exponentially decaying part $\sim \exp (-\sum_{\stackrel{i,j=0,\ua,\da}{i\neq j}} W_{ij}\, t/2)$ and the stationary solution
\be
\rho_i^{\rm stat} = (W_{ij}W_{ik} +  W_{ij}W_{jk} + W_{kj}W_{ik})
/N,
\label{eq:statsol}
\ee
with $N\equiv W_{0\ua}(W_{0\da}+W_{\ua \da}+W_{\da 0}) + W_{0\da}(W_{\ua 0}+W_{\da \ua}) +
(W_{\ua 0}+W_{\da 0})(W_{\ua \da}+W_{\da \ua})$. The number of transferred electrons per cycle is now obtained by considering the stationary solution (\ref{eq:statsol}) in each stage of the pumping cycle depicted in Fig.~\ref{fig:cycle1}. $n_{\ua,R}$ then for example becomes~\cite{tunnelrates}
\be
n_{\ua,R} = \frac{W_{\ua 0, L} W_{0\ua, R} W_{\ua \da}^2}{\prod_{B=L,R} (W_{0\ua,B}W_{\ua \da} + W_{\ua 0,B}(W_{\ua \da} + W_{\da \ua}))}.
\label{eq:statlim}
\ee
In the limit $W_{0\ua,L} \ll W_{\ua 0,L}$, $W_{\ua 0,R}\ll W_{0\ua,R}$ and $W_{\da \ua} \ll W_{\ua \da}$ for B$>$0, Eq.~(\ref{eq:statlim}) reduces to $n_{\ua,R}=1$. 

In order to check these limits and estimate $T_{\rm ad}$ and $\tau$ we use the GaAs parameters $m^{*}=0.61\cdot 10^{-31}$ kg, $g^{*}=-0.44$ and $A=90\, \mu$eV~\cite{page77}, an applied magnetic field B\,=\,8\,T and $N_{\rm eff} = \sqrt{N}=10^3$ nuclei~\cite{nuclei}. From Eq.~(\ref{eq:Tad}) we then find $T_{\rm ad} \gg 10^{-5}$ s. For typical tunneling rates $\Gam_{\rm tunnel} \sim 1\,$MHz~\cite{raising}, so that $\tau_{\rm tunnel} \ll T_{\rm ad}$ we thus obtain, assuming $\tau_{\rm switch} \leq \tau_{\rm tunnel}$~\cite{switching} a cycle duration of $\tau = 4\cdot 10^{-6}$ s.
Since $T_1=0.85$ ms at B=8 T~\cite{elze04}, we obtain for typical temperatures $T=10$ mK that $W_{\da \ua} \ll W_{\ua \da} = 1.2\cdot 10^3\, s^{-1} \ll \Gam_{\rm tunnel}$, so that $W_{\da \ua}$ can safely be neglected in (\ref{eq:statlim}). We then find $e n_{\ua,R}/\tau \sim$ 40 fA, $I_c$ = 0 A and $I_s = 0.5\cdot 10^6$ spins s$^{-1}$. 

Finally, some remarks on co-tunneling, which could also destroy the pure spin pump effect. Since no bias voltage is applied, the mechanism of inelastic co-tunneling is not allowed~\cite{defr02}. The probability of elastic co-tunneling $\sim \Gam_{\rm tunnel}^2$ and is thus strongly suppressed for high tunnel barriers. We find $I_{\rm cot} \sim \frac{e}{\tau_{\rm cot}} \sim 0.1$ fA $\ll \frac{e}{\tau}$, which agrees with recent experimental findings~\cite{hans03} and should have negligible effect on the operation of the spin pump. 

\begin{figure}
\centerline{\epsfig{figure=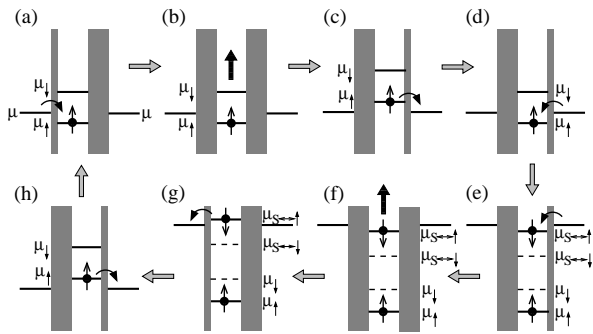,width=0.90\hsize}}
\caption[]{Spin pump turnstile with an oscillating gate voltage. The pump cycle is explained in the text.
}
\label{fig:cycle2}
\end{figure}

Instead of by switching the magnetic field, a spin-without-charge current can also be obtained in this quantum dot turnstile by using a time-dependent gate voltage. The corresponding pumping cycle is depicted in Fig.~\ref{fig:cycle2}. The first part of the cycle is very similar to Fig.~\ref{fig:cycle1}: a spin-$\ua$ electron is transferred from the left lead to the right one [(a)-(c)], only here a gate voltage is applied to shift both levels simultaneously upwards [(b)]. After the spin-$\ua$ has tunneled out, however, the rest of the cycle is different: first the levels are lowered again and a spin-$\ua$ electron tunnels in from the right lead [(d)]. The energy levels are then further lowered until the next available level $\mu_{S \leftrightarrow \ua} < \mu$, and a spin-$\da$ tunnels into the dot as well [(e)]. The dot is now occupied by a singlet and the right barrier is closed. Next, the energy levels are brought up [(f)] until $\mu_{S \leftrightarrow \ua} > \mu > \mu_{\ua}$ and the left barrier is opened. The spin-$\da$ tunnels out into the left lead [(g)], after which the left barrier is closed, the levels are brought up further until $\mu_{\ua} > \mu$, the right barrier is opened and the spin-$\ua$ electron tunnels out into the right lead [(h)]. By closing the right barrier, bringing the levels down again and opening the left barrier, a new cycle starts. Note that the spin-$\ua$ electron which enters the dot from the right lead [(d)] does not contribute to the current, but only facilitates the transfer of the spin-$\da$ electron to the left.

This pumping cycle is thus less efficient than the previous one as it requires 6 tunneling events instead of 4 to transfer two electrons. For faithful transfer of a spin-$\da$ to the left, the levels have to be lowered sufficiently slowly in (d) that a spin-$\ua$ tunnels in first. So the level position $\mu_{\da} > \mu > \mu_{\ua}$ has to be kept during at least $\tau_{\rm tunnel}$, otherwise spin-$\ua$ and spin-$\da$ have equal probabilities to tunnel in first, which results in a 50 \% probability of zero net charge and spin to be transferred.

In the ideal case, if the time required for shifting the levels is negligible compared to $\tau_{\rm tunnel}$, $\tau = 6\, \tau_{\rm tunnel}$. Using similar arguments as before~\cite{risk} we then find, assuming that in each cycle a spin-$\ua$ is transferred to the right and a spin-$\da$ to the left, $I_c=0$ A and $I_s = 0.3\cdot 10^6$ spins $s^{-1}$. Note that the applied magnetic field has to be such that the singlet state is the ground state for 2 electrons on the dot. In parallel magnetic fields the singlet state is the ground-state up to fields of at least B=12 T~\cite{hans03}, while for perpendicularly applied fields the singlet-triplet transition already occurs at B$\sim$1.3 T~\cite{kyri02}, so it is advantageous to use a parallel field in this scheme. Also, an important requirement is that the singlet-triplet spin flip time $T_{ST}$ must be much longer than the tunneling time. $T_{ST}$ had been estimated in Ref.~\cite{erli01} and was predicted to be $\sim 10^2\, {\rm s} \gg \tau_{\rm tunnel}$. Compared to the previous model, the advantage of this scheme is that pulsed gate voltages that provide fast shifting ($\leq 1$ ns) of dot levels are already available~\cite{raising}. The disadvantage is the lesser efficiency, since this second scheme requires an extra spin-$\ua$ electron to assist in the pumping proces which does not contribute to the current.

In the basic models discussed above, phase coherence plays no role. It is interesting, however, to ask whether one could also transfer the electrons in such a way that their phase information - which is important if they e.g. form part of an entangled pair - is retained. In order to do so, the switching of the magnetic field or the shifting of the energy levels has to be sufficiently fast (``sudden'') for the phase factor of the wavefunction to remain unchanged. For the oscillating magnetic field we obtain the condition, using (\ref{eq:Ham}) and following~\cite{mess58},
\be
T_{\rm rapid} \ll \frac{2 \sqrt{3} \hbar}{(\frac{7}{15} \frac{\hbar^2 \omega_c^4}{4 \omega_0^2 + \omega_c^2} + E_Z^2)^{1/2}} \approx \frac{2\sqrt{6} \sqrt{4\omega_0^2 + \omega_c^2}}{\omega_c^2}
\label{eq:rapid}
\ee
and for the oscillating gate voltage we find $T_{\rm rapid} \ll \hbar/\Del \mu$, where $\Del \mu$ is the distance over which the levels are raised or lowered. For B=8T Eq.~(\ref{eq:rapid}) yields $T_{\rm rapid} \ll 1\cdot 10^{-12}$ s, which is much faster than switching times achieved so far. For the second model, on the other hand, $T_{\rm rapid}$ is accessible for small $\Del \mu$ : if $\Del \mu < 10^{-24}$ J, we obtain $T_{\rm rapid} < 10^{-10}$ s which is within reach of sub-ns pulse switching times.

In conclusion, we have presented two schemes for realizing a spin pump quantum dot turnstile, i.e. a device that is capable of pumping quantized amounts of spin in the absence of a charge current (spin battery). In both cases the pumped spin current is quantized in units of the inverse cycle time $\tau_{\rm cycle}^{-1}$, which can be externally controlled. This turnstile is the spin analogue of the classical charge quantum dot turnstile~\cite{kouw92}. It can be used as a fully controllable double-sided bipolar spin filter, since it can be operated to either transmit spin-$\ua$ to the right and spin-$\da$ to the left or the reverse. For the same reason, it may also be used for controlled and ``on demand'' double-sided spin injection. An interesting direction for future research is to assess the fidelity of the quantized spin current, along similar lines as for quantized charge currents~\cite{kell98} and study the crossover from quantum (as in Ref.~\cite{mucc02}) spin pumping to classical spin pumping in quantum dots.

We thank R. Hanson and I.T. Vink for valuable discussions. This work has been supported by the Stichting voor Fundamenteel Onderzoek der Materie (FOM), and by the EU's Human Potential Research Network under contract No. HPRN-CT-2002-00309 (``QUACS'').


\begin{thebibliography}{99}
\bibitem{tunnelrates}
Here we have assumed that the tunnel rate through closed barriers is zero. 
Note that $\mu_{\ua} - \mu > k_B T$ when the $\ua$-spin tunnels out 
of the dot in Fig.~\ref{fig:cycle1}(c) to prevent another $\ua$-spin 
from tunneling into the dot from the right lead, and similarly for 
$\mu_\da - \mu > k_B T$ in Fig.~\ref{fig:cycle1}(f).
\bibitem{switching}
State-of-the-art techniques do not allow yet for the fast switching times of a magnetic field ($\sim$ MHz) that we require here. But since new techniques (e.g. for performing fast ESR in quantum dots) are currently being developed, faster switching times may well come within reach in the near future.
\bibitem{risk}
Note that in this second pumping scheme each electron remains in its changing energylevel during the voltage pulse, since all levels are raised or lowered by the same amount.
\end{thebibliography}
\end{document}